\documentclass{article}
\usepackage{amssymb}

\usepackage{graphicx}
\usepackage{amsmath}

\input{tcilatex}

\begin{document}

\title{Nature of the quantum insulator to superconductor and superconductor to
normal state transitions in cuprate superconductors}
\author{T. Schneider \\
Physik-Institut der Universit\"{a}t Z\"{u}rich, Winterthurerstrasse 190,\\
CH-8057 Z\"{u}rich, Switzerland}
\date{}
\maketitle

\begin{abstract}
Using the scaling theory of quantum critical phenomena we present a detailed
analysis of various superconducting properties of La$_{2-x}$Sr$_{x}$CuO$_{4}$%
, Y$_{0.8}$Ca$_{0.2}$-123, Tl-1212 and Tl-2201. Our results include: the
experimental data are fully consistent with a 2D insulator to superconductor
(ISQ) transition in the underdoped limit. The quantum critical properties
appear to be equivalent to those of a dirty 2D bosonic system with long
range Coulomb interaction. Together with previous evidence, this transition
appears to be a generic feature of cuprate superconductors. The 3D
superconductor to normal state quantum transition (SNQ) turns out to be less
unique. More extended experimental data is needed to settle the nature of
this transition unambiguously.
\end{abstract}

\section{Introduction}

A characteristic property of cuprate superconductors is the existence of a
chemical composition that gives a maximum transition temperature $T_{c}$,
separating the so-called underdoped and overdoped regimes. This behavior
appears to be a generic feature. In practice, however, there are only a few
compounds for which the composition can be varied continuously throughout
the entire doping range. La$_{2-x}$Sr$_{x}$CuO$_{4}$ is an example in which
both, the underdoped and overdoped regimes, can be accessed by variation of
the dopant concentration $x$. In the underdoped limit ( $x=x_{u}\approx 0.05$
) and $T=0$ an insulator to superconductor quantum phase (ISQ) transition
occurs, while in the overdoped limit ($x=x_{o}\approx 0.27$) a
superconductor to normal state quantum (SNQ) transition takes place \cite
{book,klosters,housten}. These quantum critical points are the endpoints of
the phase transition line $T_{c}\left( x\right) $ which adopts its maximum
around $x\approx 0.16$. Another generic feature of cuprate superconductors
emerges from the doping dependence of the anisotropy parameter $\gamma =%
\sqrt{M_{c}/M_{ab}}$, where $M_{c}$ and $M_{ab}$ are the effective pair
masses along the crystallographic c-axis and in the ab-plane, respectively.
Magnetic torque, penetration depth and resistivity measurements revealed,
that whenever the underdoped regime can be accessed, $\gamma \left( x\right) 
$ tends to diverge in the underdoped limit and decreases monotonously in the
overdoped regime \cite{book,klosters,housten,hoferG,panagopoulos}. These
behavior uncovers a dimensional crossover from 2D to 3D with increasing
dopant concentration and implies a 2D-ISQ transition in the underdoped limit.

The main purpose of this work is to explore the nature of the SNQ transition
in cuprate superconductors. Invoking the scaling theory of quantum critical
phenomena we present a detailed analysis of various superconducting
properties of La$_{2-x}$Sr$_{x}$CuO$_{4}$, Y$_{0.8}$Ca$_{0.2}$-123, Tl-1212
and Tl-2201. In La$_{2-x}$Sr$_{x}$CuO$_{4}$ these properties include the
doping dependence of\ $T_{c}$, zero temperature penetration depths \cite
{panagopoulos}, linear $T$ coefficient of specific heat coefficient \cite
{loram,momono} and in-plane penetration depth $\lambda _{ab}$ \cite
{panagopoulos}. In Y$_{0.8}$Ca$_{0.2}$-123, Tl-1212 and Tl-2201 we
concentrate on the doping dependence of $T_{c}$ and $\lambda
_{ab}^{-2}\left( T\right) $ \cite{niedermayer,bernhard}. Our main results
include: The experimental data is consistent with a 2D-ISQ transition in the
underdoped limit. Together with the previous evidence, this transition
appears to be a generic feature of cuprate superconductors \cite
{book,klosters,housten}, although in practice, there are only a few
compounds for which the dopant concentration can be varied continuously
throughout the entire doping range. Its quantum critical properties fall
into the universality class of a 2D bosonic system with long range Coulomb
interaction \cite{fisher,herbut}. The 3D-SNQ transition turns out to be more
complex. The experimental data for overdoped La$_{2-x}$Sr$_{x}$CuO$_{4}$
reveals remarkable consistency with an intermediate mean-field clean d-wave
superconductor to normal state quantum transition (CSNQ). This behavior is
not confirmed by the data for overdoped thin films \cite{locquet} and the $%
\mu $SR data for bulk Y$_{0.8}$Ca$_{0.2}$-123, Tl-1212 and Tl-220, \cite
{niedermayer,bernhard}, where disorder appears to be relevant. We explore
the applicability of a fluctuation dominated 3D-SNQ- and a mean-field d-wave
disordered superconductor to the normal state quantum (DSNQ) transition.\
Unfortunately, more extended $\mu $SR data are needed to discriminate
between these options unambiguously. When overdoped cuprate superconductors
exhibit in the normal state Fermi liquid like properties and, doping and
disorder are inseparable, the occurrence of the DSNQ transition turns out to
be more likely. In this case, doping does not control the mobile carrier
density only, but in terms of disorder and of the doping of the anisotropy
parameter $\gamma $, the nature and the dimensionality of the SNQ
transition, as well. It is shown that the doping dependence of $d/dT\left(
1/\lambda _{ab}^{2}\left( T\right) \right) $ in overdoped cuprate
superconductors allows to discriminate between the crossover from the 2D-ISQ
transition to intermediate mean-field CSNQ or to 3D -SNQ critical behavior.
It turns out, however, that more extended experimental data is needed to
settle the nature of the 3D-SNQ transition unambiguously.

The rest of the paper is organized as follows. In Sec.2 we sketch the
application of scaling theory of quantum critical phenomena to the 2D-ISQ
and SNQ transitions in cuprate superconductors. Sec.2.1 treats the 2D-ISQ
transition, Sec.2.2 is devoted to the fluctuation dominated 3D-SNQ
transition, Sec.2.3 to the mean-field d-wave clean-superconductor to normal
state quantum transition (CSNQ) and Sec.2.4 to the mean-field, d-wave,
disordered superconductor to normal state quantum transition (DSNQ). In
Sec.3 we present and discuss the analysis of the experimental data.

\section{Sketch of the scaling theory of quantum critical phenomena}

\bigskip

\subsection{Quantum insulator to superconductor transition (ISQ)}

There is accumulating evidence that cuprate superconductors undergo in the
underdoped limit and at $T=0$ a quantum insulator to superconductor (ISQ)
transition \cite{book,klosters,housten}. The transition is driven by
variation of the dopant concentration $x$ and occurs in the so called
underdoped limit $x_{u}$. Close to criticality $\left( x\gtrapprox
x_{u}\right) $ the scaling theory of quantum critical phenomena predicts
that in 2D transition temperature $T_{c}$ and in-plane penetration depth $%
\lambda _{ab}(T=0)$ scale as \cite{book,klosters,kim}: 
\begin{equation}
T_{c}=a_{ISQ}\left( x-x_{u}\right) ^{z\overline{\nu }},\frac{1}{\lambda
_{ab}^{2}(T=0)}=b_{ISQ}\left( x-x_{u}\right) ^{z\overline{\nu }}  \label{eq1}
\end{equation}
$z$ is the dynamic critical exponent of the ISQ -transition and $\overline{%
\nu }$ the exponent of the zero temperature in-plane correlation length
which diverges as $\xi _{ab}\left( T=0\right) \propto \left( x-x_{u}\right)
^{-\overline{\nu }}$. Thus in 2D $\ T_{c}$, $1/\lambda _{ab}^{2}(T=0)$ and $%
\rho _{0}$ are related by \cite{book,kim},

\begin{equation}
k_{B}T_{c}=c\left( \frac{\rho _{0c}-\rho _{0}}{\rho _{0c}}\right) ^{z\ 
\overline{\nu }}={\frac{\Phi _{0}^{2}}{16\pi ^{3}\overline{Q}_{2}}}{\frac{%
d_{s}}{\lambda _{ab}^{2}(T=0)}.}  \label{eq2}
\end{equation}
The critical residual resistivity $\rho _{0c}$ separates the insulating from
the superconducting phase. $a_{ISQ}$ and $b_{ISQ}$ are nonuniversal critical
amplitudes related by Eq.(\ref{eq2}). $d_{s}$ denotes the thickness of the
superconducting slab and $\overline{Q}_{2}$ is a universal constant.

Along the phase transition line $T_{c}\left( x\right) $, however, thermal
fluctuations dominate the finite temperature critical behavior. There is
accumulating evidence that it falls into the $3D-XY$ universality class \cite
{book}. In this case ( see Appendix A ), the transition temperature and
finite temperature critical amplitudes of in - plane correlation length $%
\left( \xi _{ab,0}^{+}\right) $\ and penetration depth $\left( \lambda
_{ab,0}\right) $ are related by $k_{B}T_{c}=\left( \Phi _{0}^{2}/\left(
16\pi ^{3}\right) \right) \mathcal{R}_{\xi }^{1/3}\xi _{ab,0}^{+}/\left(
\gamma \lambda _{ab,0}^{2}\right) $ (Eq.(\ref{eqA7})). $\gamma =\sqrt{%
M_{c}/M_{ab}}$ is a measure for the anisotropy. $M_{ab}$ and $M_{c}$ denote
the in-plane and out of plane effective pair mass, entering the spatial
gradient terms of the Ginzburg-Landau action. Supposing that there is a
2D-ISQ transition at a critical endpoint of the $3D-XY$ critical line,
matching with the quantum behavior $T_{c}{\propto }\delta ^{z\overline{\nu }}
$, $\lambda _{ab,0}^{-2}\propto {\lambda _{ab}^{-2}(0)\propto }\delta ^{z%
\overline{\nu }} $(\ref{eq1}) and $\xi _{ab,0}^{-}\propto \xi _{ab}\left(
T=0\right) \propto \delta ^{-\overline{\nu }}$ requires that (Eq.(\ref{eqA8}%
) 
\begin{equation}
\gamma =\sqrt{\frac{M_{c}}{M_{ab}}}\propto \left( x-x_{u}\right) ^{-%
\overline{\nu }}  \label{eq3}
\end{equation}
Noting that $\gamma =\infty $ corresponds to a $2D$ system of thickness $%
d_{s}$, this divergence is simply reflects the 3D-2D crossover. As
underdoped cuprate superconductors are concerned, a rapid increase of $%
\gamma \left( x\right) $ is experimentally well established in terms of
magnetic torque \cite{hoferG}, penetration depth \cite{panagopoulos} and
transport measurements. Thus, there is accumulating evidence that the
experimental data, analyzed in terms of Eqs.(\ref{eq1}), (\ref{eq2}) and (%
\ref{eq3}) are remarkably consistent with a generic 2D-ISQ transition with 
\cite{book,klosters,housten} 
\begin{equation}
z\ \overline{\nu }\approx 1,\ z=1,\ \overline{\nu }\approx 1  \label{eq4}
\end{equation}
These estimates are close to theoretical predictions \cite{fisher,herbut},
from which $z=1$ would be expected for a dirty 2D bosonic system with long
range Coulomb interactions independent of dimensionality and $\overline{\nu }%
\geq 1\approx 1.03$ . In this transition, the loss of phase coherence is due
to localization of Cooper pairs which is ultimately responsible for the ISQ
transition. Accordingly, the 2D-ISQ\ transition in heavily underdoped
cuprate superconductors appears to fall in the same universality class as
the onset of superfluidity in $^{4}$He in disordered media, corrected for
the long-range Coulomb interaction.

Noting that the singular part of the free energy density scales close to the
ISQ transition as \cite{book,kim} 
\begin{equation}
f_{s}\propto \left( x-x_{u}\right) ^{\overline{\nu }\left( D+z\right) }%
\mathcal{F}\left( T\left( x-x_{u}\right) ^{-z\overline{\nu }}\right) ,
\label{eq5}
\end{equation}
where $\mathcal{F}$ $\left( y\right) $ is a scaling function of its
argument, we obtain for the specific heat coefficient in 2D the relations, 
\begin{equation}
\gamma _{c}=\frac{C}{T}\propto \frac{d^{2}f_{s}}{dT^{2}}\propto \left(
x-x_{u}\right) ^{\overline{\nu }\left( D-z\right) },\ \ \ \frac{d\gamma _{c}%
}{dT}\propto \left( x-x_{u}\right) ^{2\overline{\nu }\left( D-2z\right) },
\label{eq6}
\end{equation}
provided that the corresponding derivatives of the scaling function at $%
y=T\left( x-x_{u}\right) =0$ exist. In the presence of a phase twist $k$ \
the free energy density reads \cite{book,kim} 
\begin{equation}
f_{s}\propto \left( x-x_{u}\right) ^{\overline{\nu }\left( D+z\right) }%
\mathcal{G}\left( k\left( x-x_{u}\right) ^{-\overline{\nu }}\right)
\label{eq7}
\end{equation}
Accordingly, the penetration depth scales as 
\begin{equation}
\frac{1}{\lambda _{ab}^{2}(T=0)}\propto \frac{d^{2}f_{s}}{dk^{2}}\propto
\left( x-x_{u}\right) ^{\overline{\nu }\left( D-2+z\right) }  \label{eq8}
\end{equation}
while in the limit $T\rightarrow 0$, 
\begin{equation}
\frac{1}{\lambda _{ab}^{2}(T)}\propto \left( x-x_{u}\right) ^{\overline{\nu }%
\left( D-2+z\right) }\mathcal{H}\left( T\left( x-x_{u}\right) ^{-z\overline{%
\nu }}\right) .  \label{eq9}
\end{equation}
holds. $\mathcal{H}\left( x\right) $ is again a scaling function of its
argument. The coefficient of the term linear in $T$ scale then as 
\begin{equation}
\frac{d}{dT}\left( \frac{1}{\lambda _{ab}^{2}}\right) \propto \left(
x-x_{u}\right) ^{\overline{\nu }\left( D-2\right) }  \label{eq10}
\end{equation}
provided that $d\mathcal{G}\left( y\right) /dy$ and $d^{2}\mathcal{G}\left(
y\right) /dy^{2}$ exist at $y=T\left( x-x_{u}\right) =0$. Combining Eqs. (%
\ref{eq6}) and (\ref{eq10}) we observe that in 2D for $z=1$ and in the
limits $T\rightarrow 0$ and $x\rightarrow x_{u}$ the relation, 
\begin{equation}
\frac{d\gamma _{c}}{dT}\propto \frac{d}{dT}\left( \frac{1}{\lambda ^{2}}%
\right) =const.,  \label{eq11}
\end{equation}
is a characteristic feature of a 2D-ISQ transition with $z=1$.

In this context we note that $\lambda _{ab}\left( 0\right) $ and $\lambda
_{c}\left( 0\right) $ are related by 
\begin{equation}
\left( \frac{\lambda _{ab}\left( 0\right) }{\lambda _{c}\left( 0\right) }%
\right) ^{2}=\frac{M_{ab}}{M_{c}}=\frac{1}{\gamma ^{2}}  \label{eq12}
\end{equation}
so that with Eqs.(\ref{eq1}) and (\ref{eq3}) 
\begin{equation}
\frac{1}{\lambda _{c}^{2}\left( 0\right) }\propto \left( x-x_{u}\right) ^{-%
\overline{\nu }\left( z+2\right) }\propto \left( x-x_{u}\right) ^{3\overline{%
\nu }}  \label{eq13}
\end{equation}
and 
\begin{equation}
T_{c}\propto \left( x-x_{u}\right) ^{z\overline{\nu }}{\propto }\left( \frac{%
1}{\lambda _{c}^{2}(0)}\right) ^{z/\left( z+2\right) }\propto \left( \frac{1%
}{\lambda _{c}^{2}(0)}\right) ^{1/3}  \label{eq14}
\end{equation}
with $z=1$. the marked difference to the relationship between $T_{c}$ and $%
1/\lambda _{ab}^{2}\left( 0\right) $ (Eq.(\ref{eq2})) simply reflects the
3D-2D crossover.

\subsection{Fluctuation dominated quantum superconductor to normal state
transition (3D-SNQ)}

As the dopant concentration increases, $\gamma $ is known to decrease
drastically \cite{book,klosters,isotope,hoferG}. Accordingly, a 2D - 3D
crossover takes place. For this reason, the SNQ occurs in 3D. As in 2D, $%
T_{c}$ and dopant concentration $x$ still scale as (Eq.(\ref{eq1}) 
\begin{equation}
T_{c}=a_{SNQ}\left( x_{o}-x\right) ^{z\overline{\nu }}  \label{eq15}
\end{equation}
In a fluctuation dominated (nonmean-field) 3D-SNQ-transition, however, the
scaling properties of the in-plane penetration depth follow from Eq.(\ref
{eq8}) so that 
\begin{equation}
\frac{1}{\lambda _{ab}^{2}\left( 0\right) }=b_{SNQ}\left( x_{o}-x\right) ^{%
\overline{\nu }\left( z+1\right) }  \label{eq16}
\end{equation}
Accordingly, transition temperature $T_{c}$ and zero temperature in-plane
penetration depth $\lambda _{ab}\left( 0\right) $ are related by 
\begin{equation}
T_{c}\propto \left( \frac{1}{\lambda _{ab}^{2}\left( 0\right) }\right) ^{%
\frac{z}{z+1}},  \label{eq17}
\end{equation}
where the factor of proportionality is nonuniversal. $x_{o}$ denotes the
dopant concentration in the overdoped limit.

From Eq.(\ref{eq5}) we obtain in 3D for the specific heat coefficient the
relations 
\begin{equation}
\gamma _{c}\propto \frac{d^{2}f}{dT^{2}}\propto \left( x_{o}-x\right) ^{%
\overline{\nu }\left( 3-z\right) },\ \ \ \frac{d\gamma _{c}}{dT}\propto
\left( x_{o}-x\right) ^{\overline{\nu }\left( 3-2z\right) },  \label{eq18}
\end{equation}
Similarly, the scaling properties of $\ 1/\lambda ^{2}(T=0)$ and $d/dT\left(
1/\lambda ^{2}(T)\right) $ are in 3D readily obtained from Eqs.(\ref{eq8}), (%
\ref{eq10}) and (\ref{eq16}). The result is 
\begin{equation}
\frac{1}{\lambda _{ab}^{2}(T=0)}=b_{SNQ}\left( x_{o}-x\right) ^{\overline{%
\nu }\left( z+1\right) },  \label{eq19}
\end{equation}
and with Eq.(\ref{eq10}) for the coefficients of the terms linear and
quadratic in $T$ , 
\begin{equation}
\left( \frac{d}{dT}\left( \frac{1}{\lambda _{ab}^{2}}\right) \right)
_{T=0}\propto \left( x_{o}-x\right) ^{\overline{\nu }}\propto \ \left( \frac{%
d\gamma _{c}}{dT}\right) _{T=0}^{1/(3-2z)}.  \label{eq20}
\end{equation}
Apparently, the term linear in $T$, considered to be a characteristic
feature of clean d-wave superconductors, vanishes here at criticality.

\subsection{Mean-field clean-superconductor to normal state quantum
transition (CSNQ)}

It appears to be well documented, however, that overdoped cuprate
superconductors exhibit in the normal state Fermi liquid like properties 
\cite{iye}. This suggests a mean-field SNQ transition in the overdoped limit.

In the clean limit $T_{c}$ and the amplitude $\Delta _{0}$ of the d-wave gap
scale as 
\begin{equation}
T_{c}=a_{SNQ}\ \left( x_{o}-x\right) ^{z\overline{\nu }}\propto \Delta _{0},
\label{eq21}
\end{equation}
where \cite{book} 
\begin{equation}
z\overline{\nu }=1/2,\ z=1,\ \overline{\nu }=1/2.  \label{eq22}
\end{equation}
While $T_{c}$ tends to zero in the overdoped regime, the mobile carrier
density in the normal state is known to remain finite \cite{iye}. Moreover,
in a clean system there are no pair breaking mechanisms. Accordingly, the
occurrence of a SNQ transition in the overdoped limit implies a vanishing
pairing interaction. In this case the zero temperature in-plane penetration
depth tends to a constant value\cite{book,singer}: 
\begin{equation}
\underset{x\rightarrow x_{o}}{\lim }\frac{1}{\lambda _{ab}^{2}\left(
x,0\right) }=const  \label{eq23}
\end{equation}
Note that in this transition, where fluctuations are absent, the scaling
relation (\ref{eq17}) does not hold. There is however the relation \cite
{makiwon} 
\begin{equation}
T_{c}\left( \frac{d}{dT}\left( \frac{\lambda _{ab}\left( 0\right) }{\lambda
_{ab}\left( T\right) }\right) ^{2}\right) _{T=0}=-0.645,  \label{eq24}
\end{equation}
between the leading temperature dependence of the in-plane penetration depth
and transition temperature. Invoking Eqs.(\ref{eq21}), (\ref{eq22}) and (\ref
{eq23}) we find, 
\begin{equation}
\left( \frac{d}{dT}\left( \frac{1}{\lambda _{ab}\left( T\right) }\right)
^{2}\right) _{T=0}=-\frac{1}{a_{SNQ}\lambda _{ab}^{2}\left( x_{o},0\right) }%
\left( x_{o}-x\right) ^{-1/2}.  \label{eq25}
\end{equation}
Moreover, in a mean-field d-wave superconductor the leading temperature
dependence of the specific heat coefficient adopts the form\cite{kubert} 
\begin{equation}
\left( T_{c}\frac{d\gamma _{c}}{dT}\right) _{T=0}=1.53\gamma _{n},\ \ \
\gamma _{n}=\frac{\pi ^{2}}{3}k_{B}^{2}N\left( 0\right) .  \label{eq26}
\end{equation}
Invoking Eqs.(\ref{eq21}) and (\ref{eq22}) we obtain 
\begin{equation}
\left( \frac{d\gamma _{c}}{dT}\right) _{T=0}=-\frac{1.53\gamma _{n}}{a_{SNQ}}%
\left( x_{o}-x\right) ^{-1/2}.  \label{eq27}
\end{equation}
Thus, close to a clean mean-field d-wave SNQ transition, the leading
temperature dependence of $1/\lambda _{ab}^{2}\left( T\right) $ and $\gamma
_{c}$ are according to Eqs.(\ref{eq25}) and (\ref{eq27}) linearly related by 
\begin{equation}
\left( \frac{d}{dT}\left( \frac{1}{\lambda _{ab}\left( T\right) }\right)
^{2}\right) _{T=0}=-\frac{1}{a_{SNQ}\lambda _{ab}^{2}\left( x_{o},0\right) }%
\left( x_{o}-x\right) ^{-1/2}\propto \left( \frac{d\gamma _{c}}{dT}\right)
_{T=0}  \label{eq28}
\end{equation}
This differs from the critical behavior at the fluctuation dominated 3D- SNQ
transition (Eq.(\ref{eq20}), where $\left( d/dT\left( 1/\lambda
_{ab}^{2}\left( T\right) \right) \right) _{T=0}$ tends to zero.

\subsection{Disordered tuned mean-field quantum superconductor to normal
state transitions (DSNQ)}

In most cuprate superconductors doping and disorder appear to be
inseparable. Moreover, in a d-wave superconductor even nonmagnetic
impurities are pair breaking and lead to a suppression of both, $T_{c}$ and $%
1/\lambda ^{2}\left( T=0\right) $ \cite{kim2}. For an s-wave impurity
potential of strength $w$ and impurity concentration $c$ the reduction of \
the mean-field $T_{c}$ \ is given by \cite{kim2} 
\begin{equation}
\ln \left( \frac{T_{c}}{T_{c0}}\right) =\Psi \left( \frac{1}{2}\right) -\Psi
\left( \frac{1}{2}+\frac{1}{4\pi rT_{c}}\right) ,  \label{eq29}
\end{equation}
where $r$ is the normal-state relaxation rate 
\begin{equation}
\frac{1}{2r}=\frac{N\left( 0\right) \pi cw^{2}}{1+\left( N\left( 0\right)
\pi cw\right) ^{2}}.  \label{eq30}
\end{equation}
There is a critical relaxation rate $r_{c}$ where $T_{c}$ vanishes. Here an
impurity tuned SNQ transition occurs. Close to the transition a mean-field
treatment yields \cite{kim2} 
\begin{equation}
T_{c}\propto \left( r-r_{c}\right) ^{z\overline{\nu }},\ z\overline{\nu }%
=1/2.  \label{eq31}
\end{equation}
Note that for small $w$ the Born approximation holds, while large $w$ , $%
w\rightarrow \infty $, corresponds to the unitary limit. In both cases, $%
T_{c}$ and $\lambda _{ab}\left( 0\right) $ scale close to the mean-field SNQ
transition as \cite{kim2} 
\begin{equation}
T_{c}\infty \left( \frac{1}{\lambda ^{2}\left( 0\right) }\right) ^{1/2}.
\label{eq32}
\end{equation}
In the unitary limit this behavior appears to hold over a rather extended
range \cite{kim2}.In contrast to the pure d-wave superconductor, the low
temperature behavior of $1/\lambda ^{2}\left( T\right) $ is no longer
linear. Below the crossover temperature $T^{\ast }$ one enters the impurity
dominated regime, where $1/\lambda ^{2}\left( T\right) \propto T^{2}$ \cite
{hirschfeld} and at somewhat higher temperatures $1/\lambda ^{2}\left(
T\right) $ displays the temperature dependence of the pure state, $1/\lambda
^{2}\left( T\right) \propto T$ \ (Eq.(\ref{eq28})). The interpolation
formula \cite{hirschfeld} 
\begin{equation}
\frac{1}{\lambda _{ab}^{2}\left( T\right) }=\frac{1}{\lambda _{ab}^{2}\left(
0\right) }+\frac{aT^{2}}{T^{\ast }+T}  \label{eq33}
\end{equation}
provides a rough measure of the crossover temperature $T^{\ast }$. It is
thus clear that in real systems where impurities cannot be avoided
impurities will dominate below $T^{\ast }$. For this reason one expects that
the nature of a mean-field SNQ transition, where $T_{c}$ tends to zero in
the overdoped regime, is dominated by disorder and, on the mean-field level,
a DSNQ transition is expected to occur.

\section{Comparison with experiment}

An essential requirement for the existence of the doping tuned 2D- ISQ and
3D-SNQ transitions is the 2D-3D crossover. Experimentally, this crossover
will manifest itself in the doping dependence of the anisotropy parameter $%
\gamma $ (Eq.(\ref{eq3})). From magnetic torque and resistivity measurements
it is known that $\gamma \left( x\right) $ tends to diverge in the
underdoped limit and decreases monotonically with increasing dopant
concentration $x$ \cite{book,klosters,hoferG}. Alternatively and even at $%
T=0 $, $\gamma $ can also be derived from penetration depth measurements in
terms of $\gamma ^{2}=\left( \lambda _{c}\left( 0\right) /\lambda
_{ab}\left( 0\right) \right) ^{2}=M_{c}/M_{ab}$ (Eq.(\ref{eq12})). In Fig.%
\ref{fig1} we show $\gamma ^{2}$ versus $x$ for La$_{2-x}$Sr$_{x}$CuO$_{4}$,
derived from the penetration depth measurements of Panagopoulos \emph{et al.}
\cite{panagopoulos}. Although the data is rather sparse in the underdoped
regime, comparison with the solid curve, confirms the doping tuned 2D-3D
crossover in the ground state. In Fig.\ref{fig2} we displayed the data of \
Panagopoulos \emph{et al.} \cite{panagopoulos}, Uemura \emph{et al.} \cite
{uemura214} and Franck \emph{et al.} \cite{franck} for La$_{2-x}$Sr$_{x}$CuO$%
_{4}$ in terms of $1/\lambda _{ab}^{2}\left( x,0\right) $ versus $x$. In the
underdoped limit ($x\rightarrow x_{u}\approx 0.05$), the data reveal
consistency with a 2D-ISQ transition, where for $z\overline{\nu }=1$ the
scaling relation $1/\lambda _{ab}^{2}\left( x,0\right) $ $\propto \left(
x-x_{u}\right) $ (Eq.(\ref{eq1})) holds. According to Fig.\ref{fig3}, this
conclusion is also supported \ by the characteristic behavior of a 2D-ISQ
transition, the linear relationship between $T_{c}$ and $1/\lambda
_{ab}^{2}\left( x,0\right) $ (Eq.(\ref{eq2})) in the underdoped regime.

In the overdoped regime, matters are much less clear. Indeed, there is just
one data point ($x=0.24$ in Fig.\ref{fig2}) favoring the suppression of $%
1/\lambda _{ab}^{2}\left( x,0\right) $ in the overdoped limit. Even from the
plot $T_{c}$ versus $1/\lambda _{ab}^{2}\left( 0\right) $ ( Fig.\ref{fig3}),
which is not affected by uncertainties in the dopant concentration, no
clear-cut conclusion can be drawn. To indicate the drastically different
behavior of \ the DSNQ and CSNQ transition, we sketched in Figs..\ref{fig2}
and \ref{fig3}, the crossover from the 2D-ISQ critical point to a 3D-SNQ\ or
mean-field DSNQ - transition in terms of dashed and solid curves, while the
dotted one mimics the crossover from the 2D-ISQ - to a CSNQ - transition,
respectively.

Since in any real cuprate superconductor, impurities and imperfections are
present, there is in a d-wave superconductor a temperature $T^{\ast }$ ( see
Eq.(\ref{eq33})), below which disorder dominates. In the SNQ transition $%
T_{c}$ tends to zero as $T_{c}=a_{SNQ}\left( x_{o}-x\right) ^{z\overline{\nu 
}}$ \ (Eq. (\ref{eq15})) and with that, there is a crossover dopant
concentration $x^{\ast }$, given by $a_{SNQ}\left( x_{o}-x^{\ast }\right) ^{z%
\overline{\nu }}=T^{\ast }$. For $x<x^{\ast }$ clean behavior will dominate,
while for $x>x^{\ast }$ disorder sets the scale. It is thus evident that
sufficiently close to the overdoped limit the critical behavior of the SNQ
transition will be dominated by disorder. Experimentally, this conclusion is
well confirmed by the in-plane penetration depth measurements on thin La$%
_{2-x}$Sr$_{x}$CuO$_{4}$ films. The data clearly reveals that in overdoped
films both, $T_{c}$ and $1/\lambda _{ab}^{2}\left( 0\right) $, are
systematically reduced $\left( 0.16<x\leq 0.23\right) $ \cite{locquet}.\
Thus, the strength of disorder has a strong influence on the crossover from
the 3D-ISQ to the SNQ transition. For nearly clean samples it will be
difficult to enter the regime where disorder controls the SNQ transition. To
illustrate this point we consider the doping dependence of $1/\lambda
_{ab}^{2}\left( T\right) $ at low temperatures. In Fig.\ref{fig5} we
displayed the data of Panagopoulos \emph{et al.} \cite{panagopoulos} for La$%
_{2-x}$Sr$_{x}$CuO$_{4}$ at various dopant concentrations. In the doping and
temperature range, $0.1\leq x\leq 0.24$ and $1\lesssim T\lesssim 10$ $K$
respectively, the data is fully consistent with a linear temperature
dependence and the magnitude of $d/dT\left( 1/\lambda _{ab}^{2}\left(
T\right) \right) $ increases with increasing dopant concentration. Although
this systematic behavior clearly uncovers d-wave superconductivity and
points to a clean mean-field CSNQ transition, a crossover to a disorder
dominated SNQ transition, setting in around $1K$, cannot be ruled out.
Indeed, fits to the interpolation formula ( Eq.(\ref{eq33}) \cite{hirschfeld}%
\ yield for the crossover temperature the estimates $T^{\ast }=1.43\pm 0.81$%
, $0.44\pm 0.0.39$, $1.43\pm 0.81$, $0.61\pm 0.16$, $1.69\pm 0.2$ and $%
0.64\pm 0.39$. for $x=0.1$, $0.15$, $0.2$, $0.22$ and $0.24$, respectively.
Noting that the data resulting from the linear fits $\left( 1/1/\lambda
_{ab}^{2}\left( 0\right) \right) $ are also included in Figs.(\ref{fig2})
and (\ref{fig3}). Consequently, the data shown in those figures stems from
rather clean samples, where the disorder dominated SNQ\ regime is pushed
rather close to the overdoped limit. To substantiate the occurrence of
intermediate clean SNQ behavior further, we turn to the doping dependence of 
$\left( d/dT\left( 1/\lambda _{ab}^{2}\left( T\right) \right) \right) _{T=0}$
\cite{panagopoulos} and $\left( d\gamma _{c}/dT\right) _{T=0}$ \cite
{loram,momono} displayed in Fig.\ref{fig6}. These properties scale linearly
over the attained doping range, approach a constant value in the underdoped
limit and their magnitude increases systematically with dopant concentration 
$x$. This is just, of what one would expect, in a crossover from the 2D-ISQ
transition with $z=1$ to a clean d-wave mean-field SNQ transition. Indeed,
close to the 2D-ISQ transition with $z=1$, $\left( d\gamma _{c}/dT\right)
_{T=0}\propto \left( d\left( 1/\lambda _{ab}^{2}\right) /dT\right)
_{T=0}=const$ holds (Eq.(\ref{eq11})), while close to the clean mean-field
SNQ transition, where Eq.(\ref{eq28}), rewritten in form 
\begin{eqnarray}
\left( \frac{d}{dT}\left( \frac{1}{\lambda _{ab}^{2}\left( T\right) }\right)
\right) _{T=0} &=&-\frac{1}{a_{SNQ}\lambda _{ab}^{2}\left( x_{o},0\right) }%
\left( x_{o}-x\right) ^{-1/2}  \notag \\
&=&-\ 0.21\left( 0.27-x\right) ^{-1/2}\ 10^{-8}\ \left( cm^{-2}K^{-1}\right)
\label{eq34}
\end{eqnarray}
applies. The solid curve in Fig.\ref{fig6} corresponds to this asymptotic
behavior with $a_{SNQ}=120.5K$ taken from $T_{c}=a_{SNQ}\left( 0.27-x\right)
^{1/2}\ \left( K\right) $ and $1/\lambda _{ab}^{2}\left( x_{0},0\right)
\approx 25.3\ \ 10^{8}$ $cm^{-2}$ derived from the data shown in Fig.\ref
{fig2}. In contrast, in the experimentally attained doping regime, the data
is inconsistent with a fluctuation dominated 3D-SNQ transition, where $%
\left( d\left( 1/\lambda _{ab}^{2}\right) /dT\right) _{T=0}\propto \left(
x_{o}-x\right) ^{\overline{\nu }}$ (Eq.(\ref{eq20})). It confirms, however,
the evidence for intermediate mean-field CSNQ behavior. In addition the
doping dependence of $\left( d\left( 1/\lambda _{ab}\left( T\right) \right)
^{2}/dT\right) _{T=0}$ turns out to be a particular suitable property to
discriminate between CSNQ and fluctuation dominated 3D-SNQ behavior.

Additional convincing evidence for a generic suppression of $T_{c}$ and $%
1/\lambda _{ab}^{2}\left( 0\right) $ in overdoped cuprates emerges from the $%
\mu $SR data displayed in Fig.\ref{fig7} for Y$_{0.8}$Ca$_{0.2}$Ba$_{2}$(Cu$%
_{1-y}$Zn$_{y}$)O$_{7-\delta }$ (Y$_{0.8}$Ca$_{0.2}$-123), Tl$_{0.5-y}$Pb$%
_{0.5+y}$Sr$_{2}$Ca$_{1-x}$Y$_{x}$Cu$_{2}$O$_{7}$(Tl-1212) \cite{bernhard}
and TlBa$_{2}$CuO$_{6+\delta }\ \left( \text{Tl-2201}\right) $ \cite
{niedermayer}. To document the reduction of \ $1/\lambda _{ab}^{2}\left(
0\right) $ as a function of the dopant concentration as well, we displayed
in Fig.\ref{fig8} $\sigma _{0}\left( 0\right) \propto 1/\lambda
_{ab}^{2}\left( 0\right) $ versus $x$ for Tl$_{0.5-y}$Pb$_{0.5+y}$Sr$_{2}$Ca$%
_{1-x}$Y$_{x}$Cu$_{2}$O$_{7}$(Tl-1212) \cite{bernhard}. Formally, this
behavior points to a generic relationship of the form 
\begin{equation}
T_{c}\propto \left( 1/\lambda _{ab}^{2}\left( 0\right) \right) ^{x}\propto
\sigma _{0}\left( 0\right) ^{x}  \label{eq35}
\end{equation}
which is inconsistent with a mean-field CSNQ, but consistent with both, a
fluctuation dominated 3D-SNQ, ( $x=z/\left( z+1\right) $, Eq.(\ref{eq17}))
or a mean-field DSNQ transition ( $x=1/2$, Eq.(\ref{eq32})).. Unfortunately,
the data shown in Figs.\ref{fig7} and \ref{fig8} are too sparse, to
discriminate between these transitions.

Since Fermi liquid like behavior appears to be a generic feature of
overdoped cuprates in the normal state, and in most cuprate superconductors
doping and disorder appear to be inseparable, we pursue the scenario of a
disorder tuned mean-field SNQ (DSNQ) transition. For this purpose, we need
the relationship between the measure of disorder and dopant concentration.
It is well documented \cite{tallon} that the empirical relation \cite
{presland}, 
\begin{equation}
T_{c}=T_{c}^{\max }\left( 1-82.6\ \left( x-0.16\right) ^{2}\right) ,
\label{eq36}
\end{equation}
describes the doping dependence of most cuprate superconductors very well. $%
T_{c}$ adopts its maximum value at $x=0.16$ and vanishes at $x_{o,\
u}=0.16\pm \sqrt{1/82.6}$ where the 2D-ISQ and 3D-SNQ transitions occur.
Close to these quantum phase transitions the doping dependence is then given
by 
\begin{equation}
T_{c}=T_{c}^{\max }\QATOPD\{ \} {2\sqrt{82.6}\left( x-x_{u}\right) }{2\sqrt{%
82.6}\left( x_{o}-x\right) }  \label{eq37}
\end{equation}
Note that $T_{c}\propto x-x_{u}$ is consistent with a 2D-ISQ transition (
see Eqs.(\ref{eq1}) and (\ref{eq4})). In the overdoped regime, where
disorder is expected to be relevant, matching with the scaling relations for
the disordered mean-field SNQ transition \ (Eqs.(\ref{eq31}) and (\ref{eq32}%
) yields 
\begin{equation}
T_{c}\propto \left( r-r_{c}\right) ^{1/2}\propto \left( x_{o}-x\right)
\label{eq38}
\end{equation}
and 
\begin{equation}
T_{c}^{2}\propto \frac{1}{\lambda _{ab}^{2}\left( 0\right) }\propto \left(
x_{o}-x\right) ^{2}.  \label{eq39}
\end{equation}
To provide a comparison with the experimental data we included in Fig.\ref
{fig8}\ the critical behavior close to the 2D-ISQ \ ( Eqs.(\ref{eq1}) and (%
\ref{eq4})) and the disorder tuned DSNQ transition \ (Eq.(\ref{eq39})) in
terms of the straight line and dashed curve, respectively. The straight line
corresponds to 
\begin{equation}
\frac{1}{\lambda _{ab}^{2}(0)}\propto \sigma _{0}\left( 0\right)
=39.12\left( x-0.05\right) ,  \label{eq40}
\end{equation}
and the dashed one to 
\begin{equation}
\frac{1}{\lambda _{ab}^{2}(0)}\propto \sigma _{0}\left( 0\right)
=5231.77\left( 0.27-x\right) ^{2}.  \label{eq41}
\end{equation}
The solid curve sketches the crossover between these quantum phase
transitions in terms of the interpolation formula 
\begin{equation}
\sigma _{0}\left( 0\right) =\left( \frac{1}{39.12\left( x-0.05\right) }+%
\frac{1}{5231.77\left( 0.27-x\right) ^{2}}\right) ^{-1}  \label{eq42}
\end{equation}
Note that the asymmetry in the doping dependence of $\sigma _{0}\left(
0\right) $ simply reflects the crossover from $1/\lambda _{ab}^{2}(0)\propto
x-x_{u}$ to $1/\lambda _{ab}^{2}(0)\propto \left( x_{o}-x\right) ^{2}$. This
differs from the symmetric doping dependence of $T_{c}$ (Eq.(\ref{eq36})).
The solid and dashed curves in Fig.\ref{fig7}, follow from Eqs.(\ref{eq36})
and (\ref{eq42}) \ with $T_{c}^{m}=106\ K$ \ for Tl$_{0.5-y}$Pb$_{0.5+y}$Sr$%
_{2}$Ca$_{1-x}$Y$_{x}$Cu$_{2}$O$_{7}$(Tl-1212). They resemble the outline of
a fly's wing\ and indicate the crossover from the 2D-ISQ transition to the
mean-field DSNQ transition. It should be recognized, however, that the
remarkable agreement between the experimental data and the interpolation
scheme, points merely to a d-wave disorder tuned SNQ transition. Indeed,
given the experimental data discussed here, it is not possible to
discriminate between the fluctuation dominated 3D-SNQ \ and the mean-field
DSNQ - transition. This conclusion is well documented in Fig.\ref{fig8}, in
terms of the missing data allowing to estimate the critical behavior close
to the SNQ transition. Here future experimental data is needed to
discriminate between the critical exponents entering Eq.(\ref{eq19}) and (%
\ref{eq39}).

In summary, we have shown that the experimental data for the doping
dependence of $T_{c}$, $\gamma $, $1/\lambda _{ab}^{2}$, $d/dT\left(
1/\lambda _{ab}^{2}\left( T\right) \right) $ and $d\gamma _{c}/dT$ , as well
as for the relation between $T_{c}$ and $1/\lambda _{ab}^{2}\left( 0\right) $
are fully consistent with a 2D-ISQ transition in underdoped La$_{2-x}$Sr$%
_{x} $CuO$_{4}$. Together with the previous evidence \cite
{book,klosters,housten}, this transition appears to be a generic feature of
cuprate superconductors, although in practice, there are only a few
compounds for which the dopant concentration can be varied continuously
throughout the entire doping range. Its quantum critical properties appear
to be equivalent to those of a dirty 2D bosonic system with long range
Coulomb interaction \cite{fisher,herbut}. This reveals that the loss of
phase coherence due to localization of Cooper pairs is responsible for the
2D-ISQ transition in cuprate superconductors.

While the experimental data for overdoped La$_{2-x}$Sr$_{x}$CuO$_{4}$
revealed remarkable consistency with an intermediate mean-field clean d-wave
superconductor to normal state quantum transition (CSNQ), this behavior is
not confirmed by the data for overdoped thin films \cite{locquet}, where
disorder is more pronounced. The behavior of these overdoped films and the $%
\mu $SR data for Y$_{0.8}$Ca$_{0.2}$-123, Tl-1212 and Tl-220 \cite
{niedermayer,bernhard} point to a rather different SNQ transition. We
studied two potential candidates:the fluctuation dominated 3D-SNQ- and the
mean-field, disordered, d-wave superconductor to the normal state quantum
(DSNQ) transition.\ Unfortunately, more extended $\mu $SR data are needed to
discriminate between these options unambiguously. When overdoped cuprate
superconductors exhibit in the normal state Fermi liquid like properties
and, doping and disorder are inseparable, the occurrence of the DSNQ
transition turned out to be more likely. In this case, doping does not
control the mobile carrier density only, but the nature of the SNQ
transition and in terms of the anisotropy parameter $\gamma $, the
dimensionality of the system as well. We identified the doping dependence of 
$d/dT\left( 1/\lambda _{ab}^{2}\left( T\right) \right) $ in overdoped
cuprate superconductors, as a property, allowing to discriminate between the
crossover from the 2D-ISQ transition to intermediate mean-field CSNQ or to
3D -SNQ critical behavior. Moreover we have shown, that the doping
dependence of \ $1/\lambda _{ab}^{2}\left( 0\right) $ also offers an
opportunity for future experimental studies to settle the nature of the
3D-SNQ transition unambiguously.

The author is indebted to C. Panagopoulos for providing the data shown in
Figs.4 and 5, and to C. Bernhard for the data displayed in Figs.7 and 8. It
is a pleasure to thank H. Keller and K.A. M\"{u}ller for useful discussions.

\begin{center}
{\Large Appendix A: Derivation of Eq.(3) }
\end{center}

There is accumulating evidence that the observed finite temperature critical
behavior of cuprate superconductors is consistent with $3D-XY$ universality 
\cite
{book,klosters,housten,hoferG,schneiariosa,schneikell,hubbard,kamal,pasler}.
This universality class is characterized by a set of critical exponents,
describing the asymptotic behavior of the correlation length $\xi _{i}^{\pm
} $, magnetic penetration depth $\lambda _{i}$, specific heat $A^{\pm }$,
etc., in terms of 
\begin{equation}
\xi _{i}^{\pm }=\xi _{i,0}^{\pm }|t|^{-\nu },\ \lambda _{i}=\lambda
_{i,0}|t|^{-\nu /2},\ C=\frac{A^{\pm }}{\alpha }|t|^{-\alpha },  \tag{A1}
\label{eqA1}
\end{equation}
where $3\nu =2-\alpha $. $i$ labels the crystallographic axes, $a,b$ and $c$%
. As usual, in the above expression $\pm $ refer to $t=T/T_{c}-1>0$ and $t<0$%
, respectively. The critical amplitudes $\xi _{i,0}^{\pm }$, $\lambda
_{i,0}^{2}$, $A^{\pm }$, etc., are nonuniversal, but there are universal
critical amplitude relations, including \cite{book,schneiariosa,schneikell} 
\begin{equation}
(k_{B}T_{c})^{3}=\left( {\frac{\Phi _{0}^{2}}{16\pi ^{3}}}\right) ^{3}{\frac{%
\xi _{x,0}^{-}\xi _{y,0}^{-}\xi _{z,0}^{-}}{\lambda _{x,0}^{2}\lambda
_{y,0}^{2}\lambda _{z,0}^{2}},}  \tag{A2}  \label{eqA2}
\end{equation}

\begin{equation}
(\mathcal{R}^{\pm })^{3}={A}^{\pm }\xi _{x,0}^{\pm }\xi _{y,0}^{\pm }\xi
_{z,0}^{\pm },  \tag{A3}  \label{eqA3}
\end{equation}
\begin{equation}
\frac{A^{+}}{A^{-}}=\mathcal{R}_{A}  \tag{A4}  \label{eqA4}
\end{equation}
and

\begin{equation}
\frac{\xi _{x,0}^{-}\xi _{y,0}^{-}\xi _{z,0}^{-}}{\xi _{x,0}^{+}\xi
_{y,0}^{+}\xi _{z,0}^{+}}=\mathcal{R}_{\xi }.  \tag{A5}  \label{eqA5}
\end{equation}
$\mathcal{R}^{\pm }$, $\mathcal{R}_{A}$ and $\mathcal{R}_{\xi }$ are
universal numbers \cite{hohenberg}. These universal relations are extensions
of the isotropic counterparts \cite{hohenberg} $\left( \xi _{x,0}^{\pm }=\xi
_{y,0}^{\pm }=\xi _{z,0}^{\pm }\text{, etc.}\right) $ to extreme type II
superconductors with effective pair mass anisotropy. This anisotropy enters
the Ginzburg-Landau action in terms of the spatial gradient terms. In this
case the correlation lengths and magnetic penetration depths scale as 
\begin{equation}
\frac{\xi _{i}^{+}}{\xi _{j}^{+}}=\sqrt{\frac{M_{j}}{M_{i}}},\ \ \frac{%
\lambda _{i}^{2}}{\lambda _{j}^{2}}=\frac{M_{i}}{M_{j}}  \tag{A6}
\label{eqA6}
\end{equation}
Since in most cuprate superconductors, $\xi _{x,0}=\xi _{y,0}=\xi _{ab,0}$, $%
\xi _{z,0}=\xi _{c,0}$, $\lambda _{x,0}^{2}=\lambda _{y,0}^{2}=\lambda
_{c,0}^{2}$ and $\lambda _{z,0}^{2}=\lambda _{c,0}^{2}$ the universal
relation (\ref{eqA2}) reduces with Eq.(\ref{eqA6}) to 
\begin{equation}
k_{B}T_{c}=\left( {\frac{\Phi _{0}^{2}}{16\pi ^{3}}}\right) {\frac{\mathcal{R%
}_{\xi }^{1/3}\xi _{ab,0}^{+}}{\lambda _{ab,0}^{2}\gamma },\ \ \gamma =}%
\sqrt{\frac{M_{c}}{M_{ab}}}  \tag{A7}  \label{eqA7}
\end{equation}
Although $T_{c}$, $\xi _{ab,0}^{+}$, $\xi _{ab,0}^{+}$ and $\gamma $ are
nonuniversal and depend on the dopant concentration, this relation holds
irrespective of the doping level along the $3D-XY$ critical line, except at
the critical endpoints, where quantum phase transitions occur. Close to the $%
2D$-ISQ transition matching with the quantum behavior requires $T_{c}\propto
\delta ^{z\overline{\nu }}$, $\lambda _{ab,0}^{-2}\propto {\lambda
_{ab}^{-2}(0)\propto }\delta ^{z\overline{\nu }}$ (Eq.(\ref{eq1})) and $\xi
_{ab,0}^{-}\propto \xi _{ab}\left( T=0\right) \propto \delta ^{-\overline{%
\nu }}$ so that 
\begin{equation}
\gamma \propto \delta ^{-\overline{\nu }}.  \tag{A8}  \label{eqA8}
\end{equation}


\FRAME{dtbpFU}{8.0309cm}{6.1747cm}{0pt}{\Qcb{}}{}{fig1.eps}{}%
Fig.1: $\gamma ^{2}=\left( \lambda _{c}\left( 0\right) /\lambda _{ab}\left(
0\right) \right) ^{2}=M_{c}/M_{ab}$ versus $x$ for La$_{2-x}$Sr$_{x}$C uO$%
_{4}$ derived from the data of \cite{panagopoulos}. The solid curve mimics
the critical behavior close to the 2D-ISQ transition according to Eqs.(\ref
{eq3}), (\ref{eq4}) and (\ref{eq16}) in terms of $\gamma ^{2}=0.79+126\
\left( x-0.05\right) ^{-2}$.\FRAME{dtbpFU}{8.0309cm}{6.1747cm}{0pt}{\Qcb{}}{%
}{fig2.eps}{}%
Fig.2: $\lambda _{ab}^{-2}\left( 0\right) $ versus $x$ for La$_{2-x}$Sr$_{x}$%
C uO$_{4}$. $\blacksquare $ : $\mu SR$ - data \cite{uemura214}; $\bullet $:
\cite{panagopoulos} ; $\blacktriangle $: \cite{franck}. The dashed line
mimics the crossover from the 2D-SIQ critical point with $z\overline{\nu }%
\approx 1$ (Eqs.(\ref{eq1}) and (\ref{eq4})) to 3D-SNQ\ or mean-field DSNQ -
criticality, while the dotted one mimics the crossover from the 2D-SIQ - to
the CSNQ - critical point (Eq.(\ref{eq23})).\FRAME{dtbpFU}{8.0309cm}{6.1747cm%
}{0pt}{\Qcb{Fig.3: $T_{c}$ versus $\protect\lambda _{ab}^{-2}\left( 0\right) 
$ for La$_{2-x}$Sr$_{x}$CuO$_{4}$. $\Delta $: \protect\cite{perret}; $%
\bigcirc $: \protect\cite{uemura214}; $\square $: \protect\cite{panagopoulos}%
. The solid and dashed curves mimic the crossover from the 2D-SIQ to the the
3D-SNQ\ or mean-field DSNQ - critical point. The dotted curve indicates the
crossover from the 2D-SIQ - to the CSNQ - critical point.}}{}{fig3.eps}{}%
\FRAME{dtbpFU}{8.0309cm}{6.1747cm}{0pt}{\Qcb{}}{}{fig4.eps}{}%
Fig.4: $T_{c}$ versus ${\lambda _{c}^{-2}(}T{=0)}$ for La$_{2-x}$Sr$_{x}$CuO$%
_{4}$.$\bigcirc $: \cite{panagopoulos}. The solid line corresponds to $T_{c}$
versus $\left( 1/{\lambda _{c}^{2}(}T{=0)}\right) ^{1/3}$ (Eq.(\ref{eq14}))
with $z=1$)\FRAME{dtbpFU}{8.0309cm}{6.1747cm}{0pt}{\Qcb{}}{}{fig5.eps}{}%
Fig.5: $\lambda _{ab}^{-2}\left( T\right) $ versus $T$ for La$_{2-x}$Sr$_{x}$%
CuO$_{4}$ at various dopant concentrations $x$. Taken from \cite
{panagopoulos}. The straight lines are linear fits yielding the estimates
for $1/\lambda _{ab}^{2}\left( 0\right) $ and $d/dT\left( 1/\lambda
_{ab}^{2}\left( 0\right) \right) $ at $T=0$, plotted in Figs.\ref{fig2} and 
\ref{fig6}, respectively.\FRAME{dtbpFU}{8.0309cm}{6.1747cm}{0pt}{\Qcb{}}{}{%
fig6.eps}{}%
Fig.6: $\left( d/dT\left( \lambda _{ab}^{-2}\left( T\right) \right) \right)
_{T=0}$ and $\left( d\gamma _{c}/dT\right) _{T=0}$ \ versus $x$ for La$_{2-x}
$Sr$_{x}$CuO$_{4}$. $\blacksquare $: Taken from \cite{panagopoulos} and $%
\bullet $: from \cite{loram,momono}. The solid curve corresponds to Eq.(\ref
{eq34}) indicating the asymptotic behavior of $\left( d/dT\left( 1/\lambda
_{ab}^{2}\left( T\right) \right) \right) _{T=0}$ close to CSNQ transition.%
\FRAME{dtbpFU}{2.5365in}{3.2759in}{0pt}{\Qcb{}}{}{fig7.eps}{}%
Fig.7: $T_{c}$ versus $\sigma _{0}\propto \lambda _{ab}^{-2}\left( 0\right) $
for Y$_{0.8}$Ca$_{0.2}$Ba$_{2}$(Cu$_{1-y}$Zn$_{y}$)O$_{7-\delta }$ (Y$_{0.8}$%
Ca$_{0.2}$-123), Tl$_{0.5-y}$Pb$_{0.5+y}$Sr$_{2}$Ca$_{1-x}$Y$_{x}$Cu$_{2}$O$%
_{7}$(Tl-1212) \cite{bernhard} and TlBa$_{2}$CuO$_{6+\delta }\ \left( \text{%
Tl-2201}\right) $ \cite{niedermayer}. The solid and dashed lines sketch the
crossover from the 2D-ISQ - \ to the mean-field DSNQ critical points in
Tl-1212 according to Eqs.(\ref{eq36}) and (\ref{eq42}).\FRAME{dtbpFU}{%
8.0309cm}{6.1747cm}{0pt}{\Qcb{}}{}{fig8.eps}{}%
Fig.8: $\sigma _{0}\propto \lambda _{ab}^{-2}\left( 0\right) $ versus hole
concentration x for Tl-1212. Experimental data taken from \cite{bernhard}.
The solid and dashed lines sketch the crossover from the 2D-QSI - \ to the
mean-field DSNQ critical points in Tl-1212 according to Eqs.(\ref{eq36}) and
(\ref{eq42}). The solid straight indicates the critical behavior close to
the 2D-ISQ critical point (Eq.(\ref{eq42})) \ and the dashed the leading
behavior close to the mean-field DSNQ critical point (Eq.(\ref{eq42})).

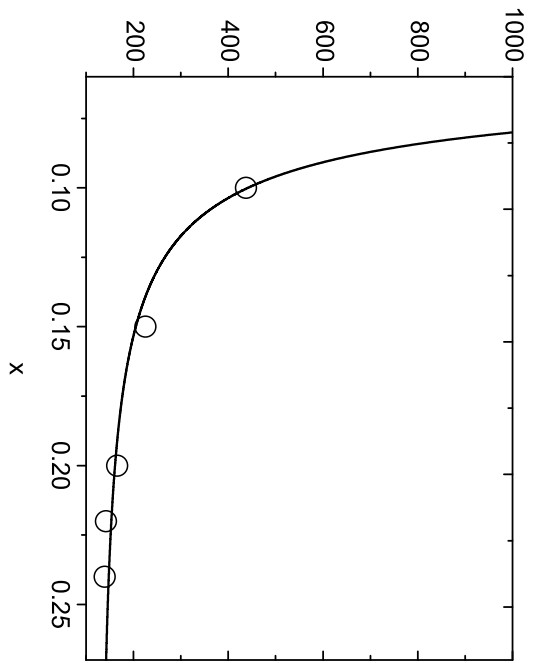
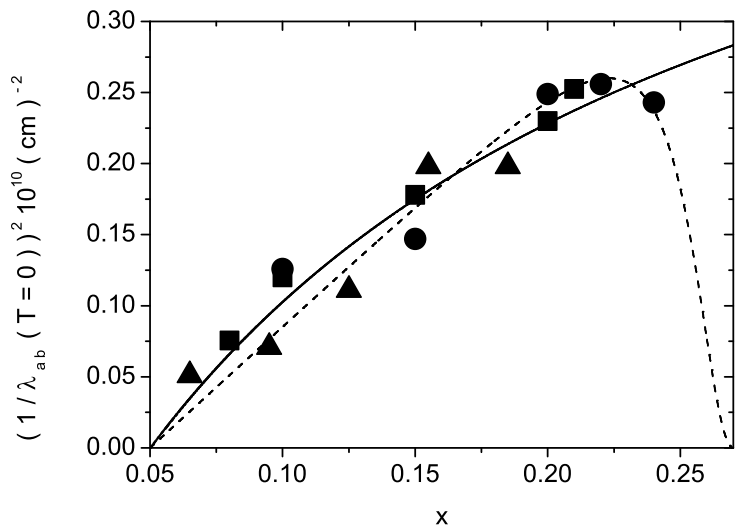
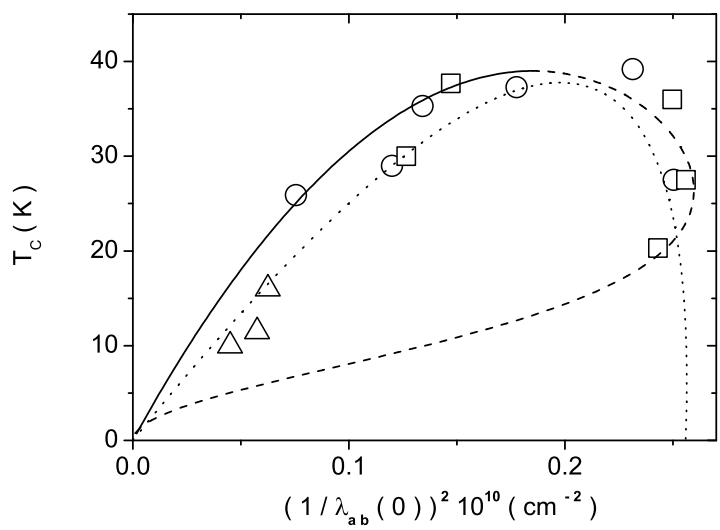
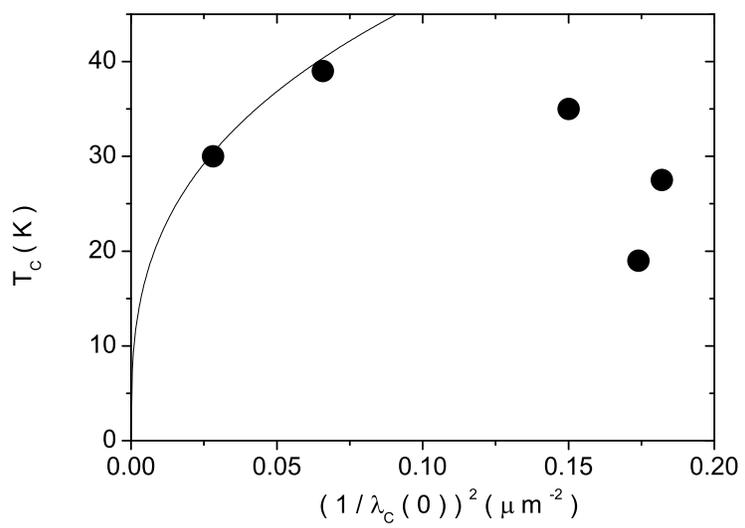
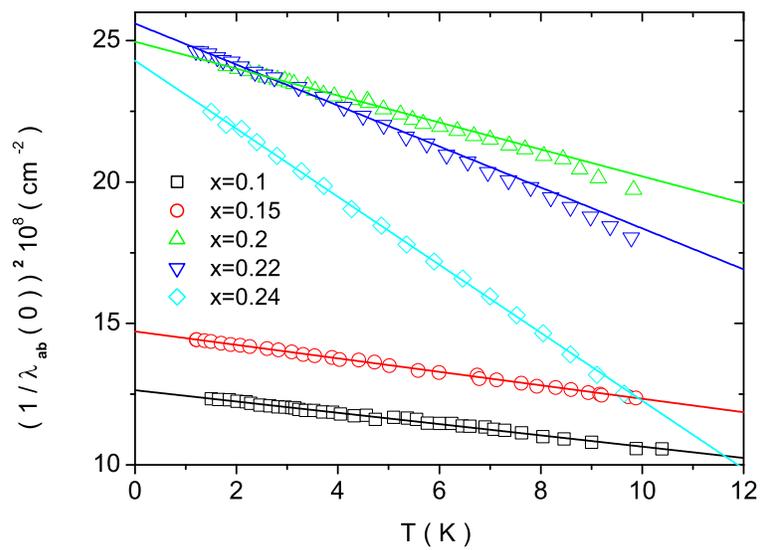
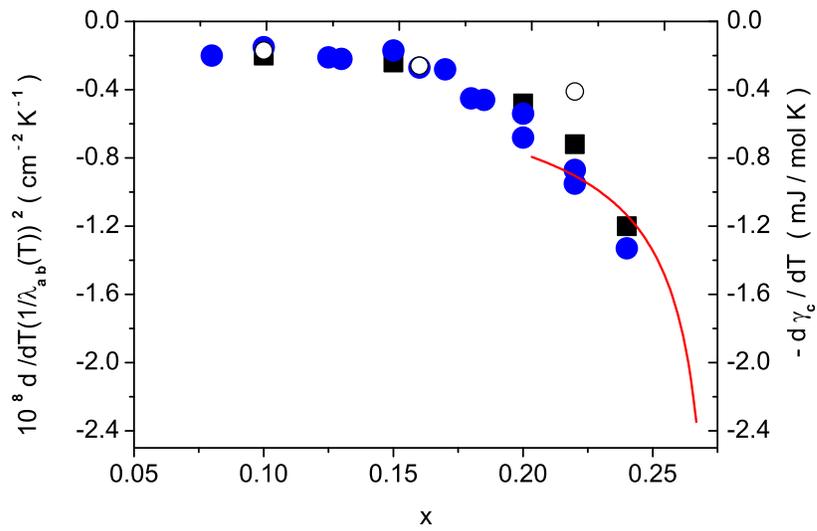
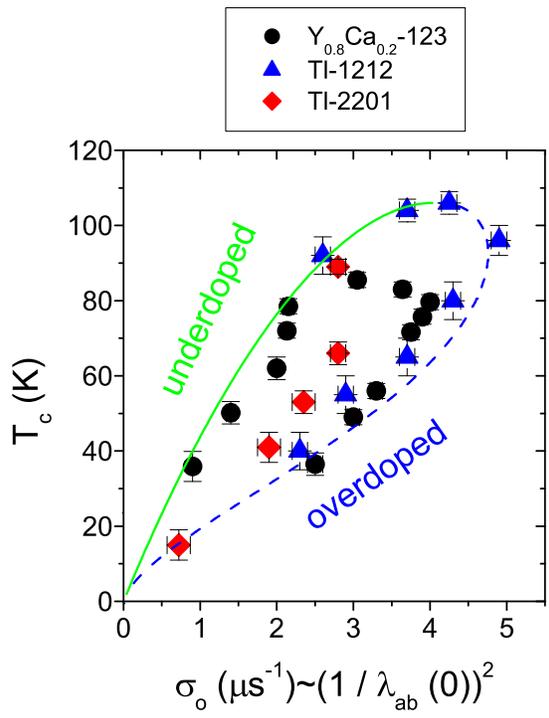
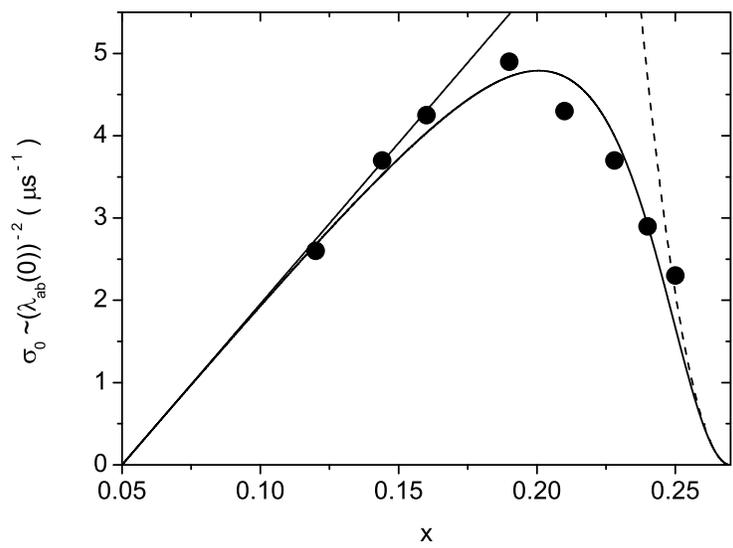
\end{document}